\documentclass[aps,prl,reprint,superscriptaddress]{revtex4-1}

\usepackage{graphicx}
\usepackage{ifpdf}
\usepackage{bm}
\usepackage{latexsym}
\usepackage{color,soul}
\usepackage{pstricks}
\usepackage{figsize}
\usepackage{float}
\usepackage{color}
\usepackage{amssymb}
\usepackage{hyperref}
\usepackage{amsmath}
\usepackage{epstopdf}
\usepackage{verbatim}
\usepackage{bbold}

\newcommand{\p}{\partial}

\begin{document}


\title{Differentiating Majorana from Andreev bound states in a superconducting circuit}



\author{Konstantin Yavilberg}
\affiliation{Department of Physics, Ben-Gurion University of the
Negev, Be’er-Sheva 84105, Israel}

\author{Eran Ginossar}
\affiliation{Advanced Technology Institute and Department of Physics, University of Surrey, Guildford GU2 7XH, United Kingdom}

\author{Eytan Grosfeld}
\affiliation{Department of Physics, Ben-Gurion University of the
Negev, Be’er-Sheva 84105, Israel}

\date{\today}

\begin{abstract}
We investigate the low-energy theory of a one-dimensional finite capacitance topological Josephson junction. Charge fluctuations across the junction couple to resonant microwave fields and can be used to probe microscopic excitations such as Majorana and Andreev bound states. This marriage between localized  microscopic degrees of freedom and macroscopic dynamics of the superconducting phase, leads to unique spectroscopic patterns which allow us to reveal the presence of Majorana fermions among the low-lying excitations.
\end{abstract}


\maketitle
{\it {Introduction}.} Quantized supercurrent oscillations in Josephson junctions strongly coupled to cavity photons, within the framework of circuit quantum electrodynamics (cQED), have become a prominent source, not only for the development of quantum processors based on the transmon 
\cite{PhysRevA.76.042319,
Majer2007,
schreier2008suppressing,
manucharyan2009fluxonium,
hassler2011top,
devoret2013superconducting,
GinossarGrosfeld2014,
PhysRevB.98.205403}, but also in the study of mesoscopic solid-state phenomena. Their high-Q superconducting resonator environment and the nonlinearity of the junction, allow precise control and high resolution microwave probing while maintaining strong coherence throughout the system. New generation hybrid devices combine additional solid-state components 
\cite{PhysRevLett.102.083602,
PhysRevLett.107.220501,
Pirkkalainen2013,
PhysRevB.92.075143,
PhysRevA.91.032307,
PhysRevLett.115.127001,
PhysRevLett.115.127002,
Tabuchi405,
Kroll2018,
Wang2019}
in order to enhance their tunability, control their responsiveness to external fields and develop a framework that can support unique quantum states, which may be difficult to probe and control in other systems.

A promising direction is to include solid state materials that when embedded inside a Josepshon junction can realize topological superconductivity via the proximity effect. Prime candidates include one-dimensional realizations of a helical liquid, including nanoribbons made of topological insulators such as Bi$_2$Se$_3$ and Bi$_2$Te$_3$ 
\cite{Zhang2009,
PhysRevB.82.045122,
doi:10.1021/nl903663a,
doi:10.1021/nl101884h,
PhysRevLett.105.136403,
PhysRevB.84.201105,
PhysRevB.86.155431}, 
or strong spin-orbit semiconductors such as InAs 
\cite{Doh272,
PhysRevLett.105.077001,
PhysRevLett.105.177002,
Alicea2011,
Chang2015}.
The resulting topological Josephson junctions	can nucleate Majorana modes whose properties can be harnessed to generate improved qubit devices
\cite{PhysRevLett.100.096407,
doi:10.1021/nl400997k,
cho2013symmetry,
PhysRevB.89.134512,
PhysRevB.94.035424,
PhysRevLett.118.126803,
PhysRevB.95.165424,
PhysRevLett.122.016801}.
So far mostly pristine topological cQED devices were theoretically studied. However, in present experimental realizations a combination of Majorana and Andreev bound states is expected to be present within the junction's weak link \cite{PhysRevB.77.184507,
PhysRevLett.103.107002,
Pillet2010,
PhysRevLett.109.056803,
PhysRevB.87.104507, 
PhysRevB.88.121109,
Ilan_2014,
van2017microwave,
PhysRevB.97.060508,
PhysRevLett.120.100502}. While their hybrid properties should play a crucial role in the development of current and future qubit devices, viable experimental methods to differentiate between their signatures in cQED are still wanting.

In this Rapid Communication we develop a methodology which allows one to study a floating mesoscopic topological Josephson junction, and to predict the experimental signatures of its low-energy excitations. The theoretical challenge stems from the interplay of the microscopic (bound states) and macroscopic (transmon) degrees of freedom controlling the dynamics of the junction. Our method identifies the relevant low-energy degrees of freedom and derives their combined dynamics. Using this method, we extract the dipole transitions of the device, which reveal the presence of bound states in the junction through a fine structure around the plasma frequency. These transitions target processes related to the Andreev bound states and their interaction with the Majorana fermions, and contain revealing signatures of these two types of bound states. 

\begin{figure}[t] 
	\centering
	\includegraphics[width=0.95\linewidth]{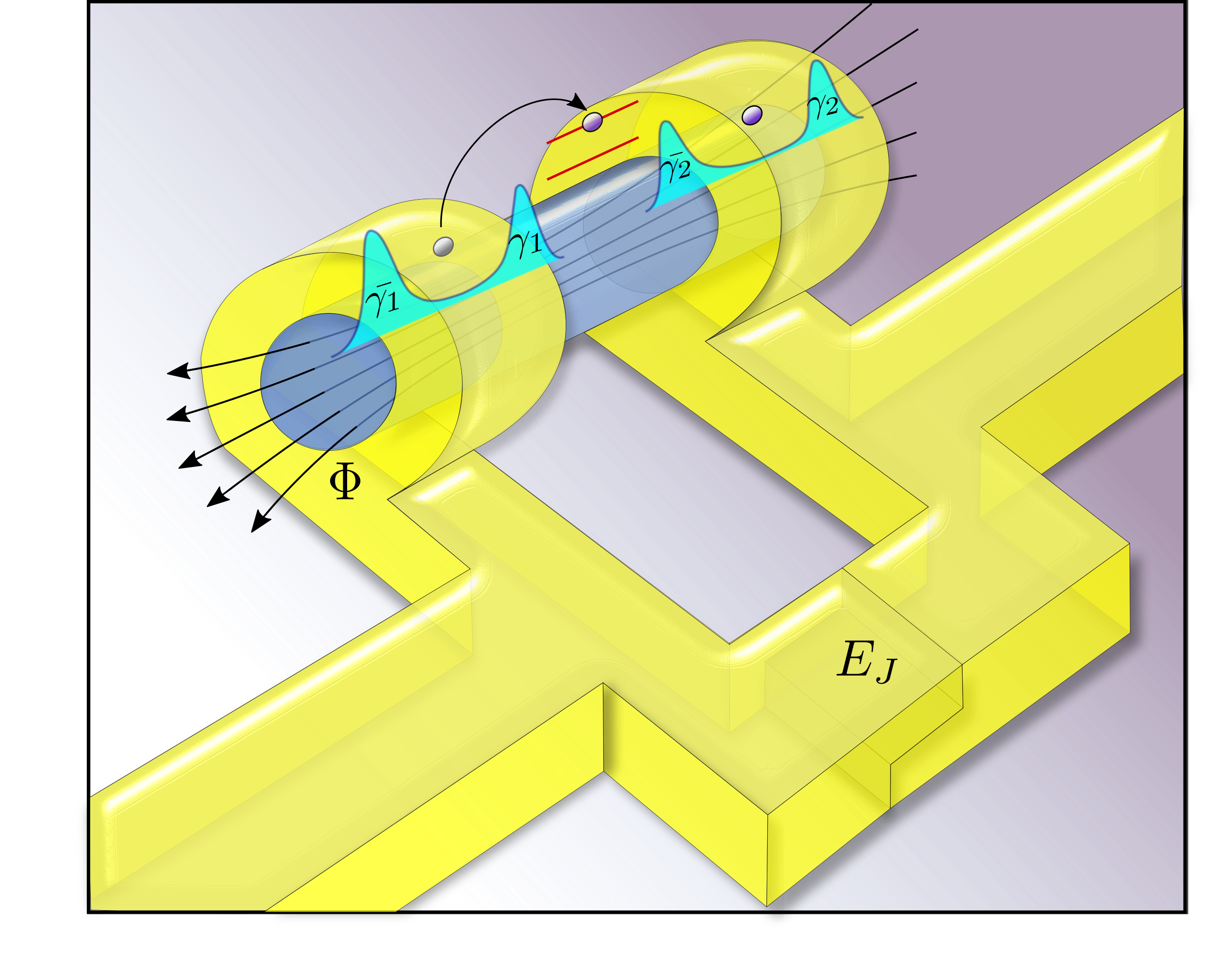}
	\caption{Schematic description of the model. A helical liquid bridging two superconducting islands nucleates Majorana and Andreev bound states. The states mediate charge transfer between the islands through single electron processes. A parallel Josephson junction generates tunneling of Cooper pairs of strength $E_J$. The helical liquid is depicted as a topological insulator nanowire threaded with a constant magnetic flux $\Phi$.}
	\label{fig0}
\end{figure}

{\it{Description of the model}.}  
We consider a one-dimensional helical liquid bridging two superconducting islands, giving rise to Majorana and Andreev bound states. For concreteness we model a topological insulator nanowire with an applied magnetic flux $\Phi$ \cite{PhysRevB.84.201105,PhysRevB.86.155431,PhysRevLett.105.136403}; we should note that our results also apply to other realizations with small modifications, such as semiconductor nanowires \cite{PhysRevLett.105.077001, PhysRevLett.105.177002}. The nanowire is connected in parallel to a regular Josephson junction with strong Josephson coupling (see Fig. \ref{fig0}). The anharmonic spectrum of the transmon, which is required for a viable qubit device, can be controlled by a side gate. Due to quantum confinement in the radial direction of the nanowire, multiple bands exist separated by $\sim v/R$, with $R$ the nanowire's radius and $v$ the Fermi velocity. We tune the magnetic flux close to $\Phi = \frac{hc}{2e}$, noting that any discrepancy from this value will open a finite magnetic gap $\Delta_B$ in the Dirac spectrum \cite{supp}, and set the chemical potential to $|\mu| < \frac{v}{R}-\Delta_B$. This ensures that only the lowest non-degenerate band is occupied, thus creating effectively a one-dimensional system. For convenience and without affecting the main results, we have set $\mu=0$ throughout.

For a complete model of the system we consider the action $S = S_{\text{sc}} + S_{_W} + S_\text{tun} + S_J$. The first term $S_{\text{sc}} = \sum_j\int dt dz \Psi_j^\dag G_j^{-1} \Psi_j$, describes the proximity-induced superconductivity in the left ($j=1$) and right ($j=2$) islands, given by the Green's function $G_j^{-1} = i\partial_t -\left(iv\partial_z \sigma_y +\Delta_B\sigma_z+\frac{\partial_t\phi_j}{2}\right)\tau_z+\Delta \sigma_y \tau_y$.  
Here $\sigma_i$ and $\tau_i$ are Pauli matrices in spin and Nambu space respectively, with the spinor $\Psi = \frac{1}{\sqrt{2}}(\psi_\uparrow, \psi_\downarrow, \psi^\dag_\uparrow, \psi^\dag_\downarrow)^T$. The superconducting phase in the pairing term $\Delta  e^{i\phi_j(t)}$ is treated beyond the mean-field approximation, which allows us to take into account the effects of charge fluctuations. In writing $S_{\text{sc}}$ we employed the gauge transformation $\Psi \rightarrow e^{[i \phi(t)/2] \tau_z} \Psi$ which removes the phase from the pairing term and adds a coupling of $\partial_t\phi_j(t)$ to the density $\Psi^\dag_j\Psi_j$.  The weak link is modeled as a two-state system, given by $S_{_W} = \int dt\; \mathcal{C}^\dag \left(i\partial_t -\varepsilon \tau_z - \Delta_B\sigma_z\tau_z\right)\mathcal{C} - U c^\dag_\uparrow c_\uparrow c^\dag_\downarrow c_\downarrow$, where $\mathcal{C}=\frac{1}{\sqrt{2}}(c_\uparrow, c_\downarrow, -c_\uparrow^\dag, -c_\downarrow^\dag)^T$ with the operators $c_\uparrow$, $c_\downarrow$ representing low-energy modes with spin orientation along the nanowire. This is justified due to the finite size of the weak link  and the resulting level quantization. We have also included $\varepsilon$ which can be controlled by a local gate operating on the weak link, and a repulsive Coulomb interaction $U$. We assume that the coupling of the weak link to the islands is given by a tunneling term of the form $S_\text{tun} = \sum_j\left(\lambda\int dt\Psi_j^\dag(0) e^{\frac{i\phi_j(t)}{2}\tau_z} \mathcal{C} + \text{h.c.}\right)$. This term can be realized by locally narrowing the nanowire near the edges of the weak link \cite{PhysRevB.95.165424}, which opens a magnetic gap and results in tunnel barriers. An alternative approach which does not require a tunnel junction is presented in \cite{supp}. The action of the parallel regular Josephson junction is given by $S_J = \int dt 
\left[\frac{(\dot{\phi}_1-\dot{\phi}_2)^2}{16E_C} + \frac{(\dot{\phi}_1+\dot{\phi}_2)^2}{16E_C'} + E_J \cos(\phi_1-\phi_2)\right]$. Here $E_C$ and $E_C'$ define the scale of the charging effect, originating from the finite capacitance of the mesoscopic device, with the ratio $E_C'/E_C \leq 1$ controlling the strength of the mutual capacitance. Throughout we will assume that the system operates in the transmon regime $E_C,E_C' \ll E_J$ \cite{PhysRevA.76.042319}, where $E_J$ is the Josephson energy. We consider the case where there is no flux penetration through the loop created by the two parallel junctions (see Fig. \ref{fig0}).
 
\begin{figure*}[t] 
	\centering
	\includegraphics[width=0.8\linewidth]{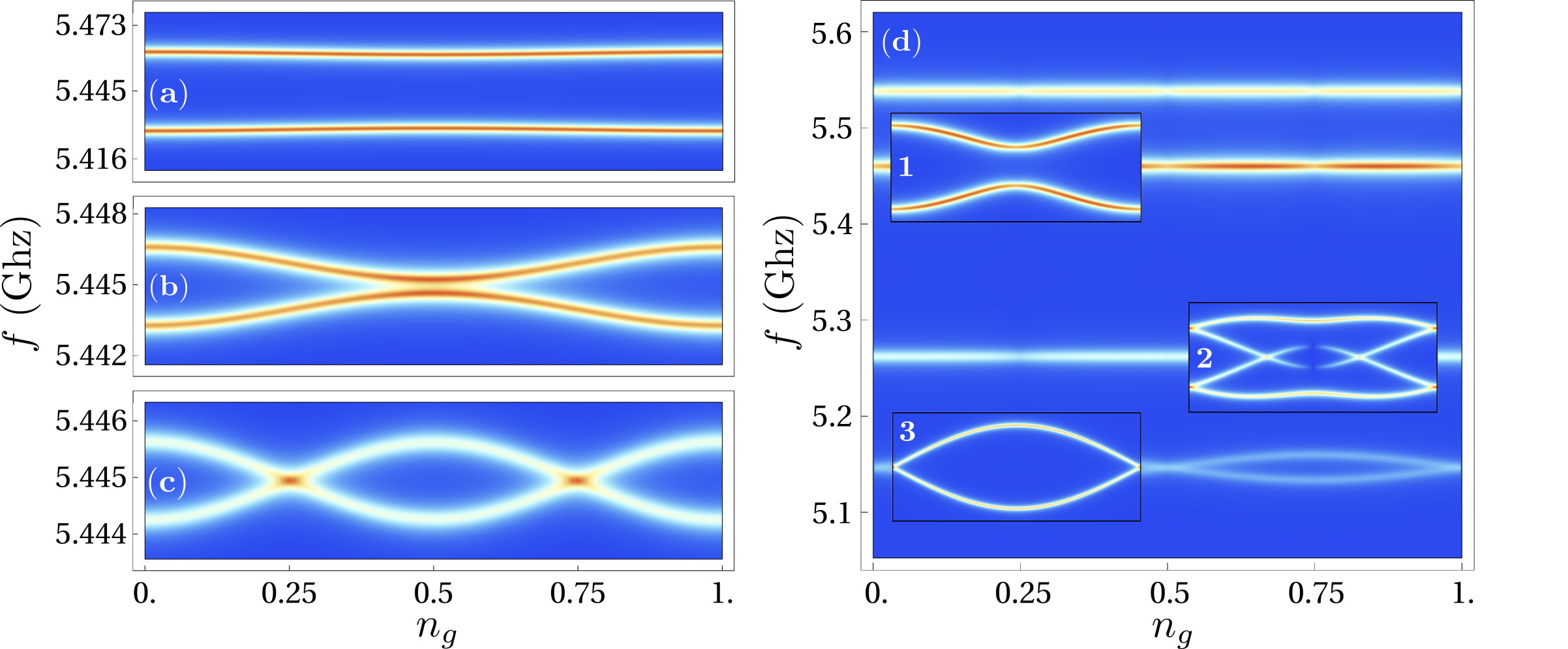}
	\caption{Predicted spectroscopic signatures for different configurations of bound states.  Dipole transition lines are presented as a function of $n_g$ for the lowest-energy sector. We use values typical for a transmon, taking $E_C/2\pi  = 0.4$ GHz, $E_J/E_C = 27$, $\Delta/E_J=10$, and $\varepsilon/E_C = 0.8$ with $U/E_C=0.6$ for the weak link parameters. We have also set $E_C'/E_C = 1$, as any deviation of this ratio from unity results in a simple renormalization of the rest of the parameters. In the case $\Delta_B=0$ we look at three different coupling strengths: (a) $\Gamma/E_C=0.5$, $\kappa/2\pi=3$ MHz, (b) $\Gamma/E_C=0.12$, $\kappa/2\pi=0.4$ MHz, and (c) $\Gamma/E_C=0.04$, $\kappa/2\pi=0.3$ MHz. All lines have the same periodicity $n_g \rightarrow n_g+1$, indicating tunneling of solely Cooper pairs. In (d) the magnetic gap is increased to $\Delta_B/E_C=11.8$, which results in additional transition lines. With the presence of Majorana fermions all transition lines exhibit an $n_g \rightarrow n_g+1/2$ periodicity due to single particle tunneling. We have set $\Gamma/E_C=0.01$, $\varepsilon/E_C = 1.3$ and $U/E_C = 2.3$. Patterns fall into three types according to their behavior at $n_g=1/4$, as seen in the magnified insets. Insets $1$ and $3$ show patterns characteristic to Andreev bound states; however, due to the mixing of fermionic parity they develop a gap at $n_g=1/4$ of size $\Delta f = 0.57$ MHz and $\Delta f = 26.4$ MHz, respectively. Inset $2$ shows the effect of Majorana fermions and Andreev bound states hybridization in its most distinctive form. The central lines show spectral holes, and are enveloped by a form of type $1$ with an increased gap $\Delta f = 1.86$ MHz. In insets $1$ and $2$ we used $\kappa/2\pi=0.1$ MHz, and in inset $3$ we used $\kappa/2\pi=0.9$ MHz.}
	\label{fig1}
\end{figure*} 
 
The dynamics of the mesoscopic topological junction is dominated by a set of degrees of freedom for which we now derive an effective theory. The theory accounts for the interaction of Cooper pairs with the bound states, by systematically integrating all highly fluctuating degrees of freedom \cite{supp}. This results in an effective Hamiltonian  $H_\text{eff} =  H_\mathcal{C} + H_\gamma + H_T$ which we later use for our main analysis of the system. Here $H_T$ is a modified transmon Hamiltonian
\begin{equation}
\label{eq: eff ham-Josephson}
\begin{aligned}
H_T \! =   4E_{C}\left(\hat{n}-n_g \right)^2  
 + E_{C}'\left(\hat{N}^2 \!+2\alpha\mathcal{C}^\dag \tau_z \mathcal{C} \hat{N}\right)
\! -E_J \cos(\hat{\varphi}),
\end{aligned}
\end{equation}
where $\hat{n}= \frac{1}{2}\left(\hat{n}_1-\hat{n}_2\right)$ is the relative number of Cooper pairs between the islands, $\hat{N} = \hat{n}_1+\hat{n}_2$ is the total number of Cooper pairs exceeding neutrality in the islands and $\hat{\varphi} = \hat{\phi}_1-\hat{\phi}_2$ is the phase difference conjugate to $\hat{n}$. The operator $e^{-i q\hat{\varphi}}$ ($e^{i q\hat{\varphi}}$) transfers a charge $q$ from the left to the right (right to the left) island. We redefined $E_C$ and $E_C'$ to include the capacitance of both the topological Josephson junction and the rest of the transmon. A side gate generates an offset charge $n_g$, measured in units of the Cooper pair charge.
The parameter $\alpha \equiv \alpha_c +\frac{\lambda^2}{v\Delta}\left(1-\frac{\Delta_B^2}{\Delta^2}\right)^{-1}$, controls the  electrostatic interaction between the weak link and the islands, and is comprised of two contributions: 
one is capacitive, given by a phenomenological parameter $\alpha_c$ which depends on the geometry of the device, and the other is a consequence of the induced superconductivity in the weak link. 

The weak link is governed by
\begin{equation}
\label{eq: eff ham-weak}
\begin{aligned}
H_\mathcal{C} & = (\tilde{\varepsilon} + \Delta_B)c^\dag_\uparrow c_\uparrow + (\tilde{\varepsilon} - \Delta_B)c^\dag_\downarrow c_\downarrow 
 + \tilde{U}c^\dag_\uparrow c_\uparrow c^\dag_\downarrow c_\downarrow\\
& + 2\Gamma \cos(\hat{\varphi}/2) 
\left(e^{i\hat{\delta}} c_\uparrow c_\downarrow  
+ e^{-i\hat{\delta}} c^\dag_\downarrow c^\dag_\uparrow \right),
\end{aligned}
\end{equation}
where $\Gamma=\lambda^2/v$ is the induced pairing and $\hat{\delta} = \frac{1}{2}(\hat{\phi}_1+\hat{\phi}_2)$ is the average phase conjugate to $\hat{N}$. The operator $e^{-i q\hat{\delta}}$ transfers a charge $q$ from the weak link to the islands. The induced pairing has emerged from the integration of high-energy degrees of freedom so it should satisfy $\Gamma\ll\Delta$. To ensure the presence of a low lying Andreev bound state we further assume $\Gamma \ll E_J$. The rest of the parameters were modified to $\tilde{U} = U + 2E_C' \alpha^2$ and $\tilde{\varepsilon} = \varepsilon + E_C' \alpha^2$. The coupling to the Majorana fermions is given by
\begin{equation}
\label{eq: eff ham-majorana}
H_\gamma = we^{i\hat{\delta}/2} \left[ie^{i\hat{\varphi}/4} 	(c_\uparrow + c_\downarrow)\gamma_1
+ e^{-i\hat{\varphi}/4} (c_\uparrow - c_\downarrow)\bar{\gamma}_2\right] + \text{h.c.},
\end{equation}
where $\gamma_1$, $\bar{\gamma}_2$ are Hermitian operators denoting the Majorana fermions localized near the weak link and $w \sim \sqrt{\Gamma \Delta_B}$. We assume negligible hybridization  
with the Majorana fermions $\bar{\gamma}_1, \gamma_2$ at the nanowire's remote ends and exclude them from the model \cite{supp}. As the parity in each island is given by the occupation of the non-local zero modes $f_1$ and $f_2$, defined by $\gamma_1 = i(f_1^\dag - f_1)$ and $\bar{\gamma}_2 = f_2^\dag + f_2$ \cite{kitaev2007unpaired}, the transfer of charge in Eq.~(\ref{eq: eff ham-majorana}) is also accompanied by a change of fermionic parity. Since the system is only capacitively shunted the total number of particles is conserved and can be fixed by a neutrality condition $ 2\hat{N} + \hat{n}_{_W} = 0$, where $\hat{n}_{_W} = c^\dag_\uparrow c_\uparrow + c^\dag_\downarrow c_\downarrow$. With this constraint and the different parity combinations, we end up with eight different subspaces denoted by $|p_1,p_2,\sigma_{_W}\rangle$, where $p_j=0,1$ indicates the occupation of $f_j$ and $\sigma_{_W}=0,\uparrow,\downarrow, \uparrow\downarrow$ correspond to the spin configurations in the weak link. 

{\it{Spectroscopic signatures of bound states.}}
To study the effect of the bound states on the spectroscopic signatures, we take for a long wire $U\sim E_C$ and $\alpha_c \ll 1$. We first consider the case where the Majorana fermions are absent, by setting $\Delta_B=0$. The resulting Hamiltonian conserves fermionic parity. By projecting $H_\text{eff}$ onto the $\{|0,0,0\rangle,|0,0,\uparrow\downarrow\rangle \}$ subspace we obtain
\begin{equation}
\label{eq: ABS - H}
\mathcal{H} = 
\begin{pmatrix}
H_{T}[n_{_W}=0] & 2\Gamma \cos(\hat{\varphi}/2)  \\
2\Gamma \cos(\hat{\varphi}/2) &   H_{T}[n_{_W}=2]+2\tilde{\varepsilon} + \tilde{U} 
\end{pmatrix},
\end{equation}
where $n_{_W} = \langle p_1,p_2,\sigma_{_W}|\hat{n}_{_W}|p_1,p_2,\sigma_{_W}\rangle$. We can get a qualitative picture by focusing on solutions with $\varphi \ll 1$, as is characteristic to the transmon regime. Ignoring the effect of the offset charge, a straightforward diagonalization of Eq. (\ref{eq: ABS - H}) gives us two independent sectors $\mathcal{H}_\pm \simeq 4E_C \hat{n}^2 +E_J \hat{\varphi}^2/2\pm \sqrt{(E_C' +2\varepsilon+U)^2 + 16\Gamma^2}/2$, corresponding to two shifted harmonic oscillators whose frequency $\omega_p = \sqrt{8E_C E_J}$ is  the plasma frequency. Higher order contributions in $\varphi$ lead to an anharmonicity of order $E_C$. The split spectrum is a result of the Andreev bound states inducing additional charge fluctuations in the weak link \cite{PhysRevB.46.12573,PhysRevLett.82.3685} as compared to the traditional transmon.
One can appreciate this by looking at the charge distribution given by $\langle \hat{n}_{_W}\rangle_\pm \simeq 1\pm \tanh\left(\frac{E_C'+2\varepsilon+U}{4\Gamma}\right)$ in each sector of $\mathcal{H}_\pm$. This was calculated using the eigenstates of Eq. (\ref{eq: ABS - H}) with $\varphi \ll 1$. To obtain a quantitative description of the model we construct the Hilbert space using the eigenstates of $\hat{n}$ and $\hat{n}_{_W}$. Since only Cooper pairs tunnel in this regime, the $n_{_W}=0$ sector imposes $n \in \mathbb{Z}$, while in the $n_{_W}=2$ sector $n \in \mathbb{Z}+1/2$, due to the absence of a Cooper pair in one of the islands. This division between the sectors is illustrated in the dependence of the energy spectrum on $n_g$ (the charge dispersion \cite{PhysRevA.76.042319,schreier2008suppressing}), and can be seen in the spectroscopic signatures (Fig. \ref{fig1}). 
\begin{figure}[t!] 
	\centering
	\includegraphics[width=1.0\linewidth]{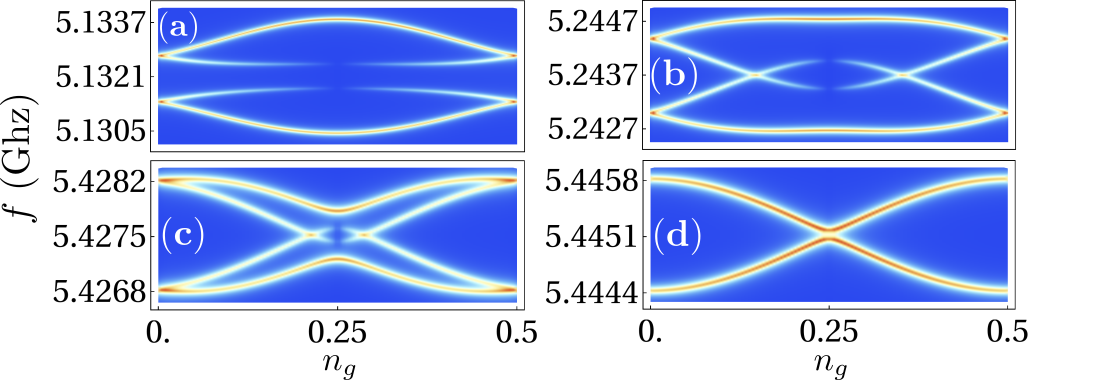}
	\caption{Dependence of the bound-states hybrid pattern on the magnetic gap. Here we show dipole transitions as a function of $n_g$, following the pattern in inset $2$ of Fig. \ref{fig1}, for the values: (a) $\Delta_B=11E_C$, (b) $\Delta_B=11.3E_C$, (c) $\Delta_B=12E_C$, and (d) $\Delta_B=12.5E_C$. The patterns have a similar dipole magnitude and are plotted with $\kappa/2\pi = 0.1$ MHz. The rest of the parameters are the same as in Fig. \ref{fig1} (d).}
	\label{fig3}
\end{figure}
The charge fluctuations between the two sides of the junction result in a coupling of the system to an electromagnetic field via the dipole moment, proportional to $\hat{n}$. The spectroscopic pattern is given by the cavity response $S_{ij}(\omega) = |\langle i|\hat{n}| j\rangle|^2 \left(\frac{\kappa^2}{[\omega - E_{ij}(n_g)]^2 + (\kappa/2)^2}\right)$, calculated as function of $n_g$. Here $\omega$ is the frequency of the photons, $\kappa$ is determined by the quality factor of the cavity and $E_{ij}(n_g)$ is the energy difference between the $i$'th and $j$'th level of $H_\text{eff}$.
As in the traditional transmon, the dominant dipole transitions are between neighboring levels separated by $\sim \omega_p$. Here, however, each sector of $\mathcal{H}_\pm$ contributes a transition line shifted with respect to its partner by $n_g \rightarrow n_g+1/2$, which results in a doublet-like pattern. For $\Gamma \rightarrow 0$ the transition lines cross at the degeneracy points $n_g=1/4+\mathbb{Z}/2$ in a manner which is seen in experimental measurements of the transmon \cite{schreier2008suppressing} and is usually a result of quasi-particle poisoning.

The dependence of the charge distribution $\langle \hat{n}_{_W}\rangle_\pm$ on the local gate $\varepsilon$ suggests a special symmetry point at $\varepsilon = -(E_C'+U)/2$. By tuning the system to this point, each sector of $\mathcal{H}_\pm$ contributes a single fermion to the weak link which occupies the Andreev bound state and results in an added pair of transition lines to Figs.~\ref{fig1} (a)-\ref{fig1}(c). This behavior is superficially similar to the Majorana-transmon \cite{GinossarGrosfeld2014,PhysRevB.92.075143}, where neighboring Majorana fermions hybridize in the weak link. 
By shifting the gate away from this finely tuned symmetry point, one can easily distinguish between the role performed by Andreev bound states and that of neutral Majorana fermions, as the latter are not affected by the gate.

We now change the flux in order to open a magnetic gap $\Delta_B\neq 0$. This uncovers the inherent differences between Andreev bound states and Majorana fermions, as observed in their distinct sensitivity to $n_g$. The two combined bound states result in a rich array of parity configurations due to single fermion transfer. The $n_{_W}=0,2$ subspaces now have variants of even and odd fermionic parity on each side while in total keeping a symmetric combination  ($p_1 = p_2$). In both variants a spin singlet is transferred between the islands without any direct response to $\Delta_B$. The $n_{_W}=1$ subspaces, on the other hand, which have an asymmetric parity combination ($p_1 \neq p_2$), accommodate spin-polarized Andreev bound states which hybridize with the Majorana fermions and result in a Zeeman splitting around the anharmonic transmon levels. The symmetric and asymmetric subspaces couple to each other with strength $\sim \sqrt{\Gamma \Delta_B}$, and due to single fermion tunneling the periodicity of the spectroscopic patterns is halved with respect to $n_g$. All the dipole transitions (see Fig. \ref{fig1}) are grouped into bands with a bandwidth determined by the charge dispersion  $\sim e^{-\sqrt{8E_J/E_C}}$ . When approaching $n_g=1/4$, the correlations between the subspaces cluster into three distinct forms. Two of the forms have transitions lines which can be distinguished by their curvature near $n_g=1/4$ and a shift of $n_g\rightarrow n_g+1/4$, one having a $\sim \cos^2(2\pi n_g)$ dependence while the other is $\sim \sin^2(2\pi n_g)$. This shift in the offset charge persists even for very small values of $\Delta_B$, and represents the difference in the energy spectrum between the symmetric and asymmetric parity  subspaces. The third form, which is characterized by the hybridization between Majorana and Andreev bound states, shows forbidden transitions near $n_g=1/4$, indicating the presence of Majorana fermions. The exact pattern is not rigid as can be seen in Fig. \ref{fig3}. By increasing $\Delta_B$ the band gradually changes its curvature and reduces its gap size, while still retaining the forbidden transitions. Thus a sweep of the magnetic flux reveals the unmistakable transition lines characteristic of Majorana fermions. Note that all three patterns can change from one form to the other, as varying the flux will inevitably create level crossings when $\Delta_B \sim \omega_p$. 

{\it{Conclusion.}}
In this work we investigated the physics of coupled low-energy bound states in a one-dimensional topological Josephson junction, where charging effects play an important role, by developing an effective theory for the physics of the weak link. We have shown that we can tune the junction between two remarkably different behaviors. The first is characterized by the absence of Majorana fermions, with Andreev bound states generating dipole transitions similar to those found in the traditional transmon. The second, in contrast, marked by the nucleation of Majorana fermions, displays a striking difference in the vicinity of the $n_g=1/4$ point, where some of the transition lines develop a vanishing intensity. The reason for this behavior is traced to a destructive interference between different parity states, mediated by the Majorana fermions. This signature emerges despite the presence of Andreev bound states and is distinct from their behavior. While zeros in the intensities at $n_g=1/4$ might occur accidentally also in the absence of Majorana fermions, the application of a local gate reveals the persistent neutrality of the Majorana fermions by maintaining a sharp zero at this value of the offset charge. Such a measurement would benefit from the unparalleled sensitivity of the cQED framework, which already accomplished experimental feats ranging from the detection of two-level defects in the oxides \cite{PhysRevLett.95.210503} to single photon detection \cite{Wallraff2004,Schuster2007} in the cavity. The same non-invasive methods can allow unprecedented accuracy for the detection of Majorana and Andreev bound states, as well as the characterization of this hybrid Majorana-Andreev-transmon model, a natural precursor for a qubit device.

\begin{acknowledgments}
{\it Acknowledgments.} This project has received funding from the European Union’s Horizon 2020 research and innovation programme under grant agreement No.~766714. K.Y. and E.Gr. acknowledge support from the Israel Science Foundation under Grant No.~1626/16.
\end{acknowledgments}

%


\newpage

\onecolumngrid
\begin{center}
  \textbf{\large Differentiating Majorana from Andreev bound states in a superconducting circuit: Supplementary Material}\\[.5cm]
  Konstantin Yavilberg,$^{1}$ Eran Ginossar,$^{2}$ and Eytan Grosfeld$^1$\\[.1cm]
  
  {\itshape ${}^1$Department of Physics, Ben-Gurion University of the
Negev, Be’er-Sheva 84105, Israel\\
  ${}^2$Advanced Technology Institute and Department of Physics,\\ University of Surrey, Guildford GU2 7XH, United Kingdom\\}
(Dated: \today)\\[1cm]
\end{center}
\twocolumngrid

\setcounter{equation}{0}
\setcounter{figure}{0}
\setcounter{table}{0}
\setcounter{page}{1}
\renewcommand{\theequation}{S\arabic{equation}}
\renewcommand{\thefigure}{S\arabic{figure}}

\section{Topological Insulator Nanowire}
\label{TIN}
Let us consider a cylindrical topological insulator with radius $R$ threaded with magnetic flux along its axis. Such systems with curved surfaces were studied previously \cite{PhysRevB.84.201105_supp,PhysRevB.79.245331_supp,PhysRevB.86.155431_supp,PhysRevLett.105.136403_supp} and we outline here only the details needed for our setup. This includes the realization of the topological insulator as a one-dimensional system and a derivation of the Majorana zero modes when superconducting pairing is added. 

The surface states of a topological insulator in cylindrical coordinates is given by the Hamiltonian
\begin{equation}
\label{eq: TIN-surface hamiltonian}
\mathcal{H}_\text{surf} = iv\partial_z\left( \sigma_y\cos \theta -\sigma_x\sin \theta\right) - \frac{v\sigma_z}{R}\left(i\partial_\theta+\Phi\right),
\end{equation}
with Fermi velocity $v$ and a dimensionless flux $\Phi$ which includes both orbital and Zeeman contributions. The reduction to a one-dimensional system can be understood more easily by rotating the Hamiltonian with
\begin{equation}
\label{eq: TIN-surface hamiltonian rotated}
e^{-\frac{i\theta\sigma_z}{2}}\mathcal{H}_\text{surf} e^{\frac{i\theta\sigma_z}{2}} = iv\partial_z \sigma_y - \frac{v\sigma_z}{R}\left(i\partial_\theta + \Phi\right).
\end{equation}
This in turn changes the angular boundary condition to $4\pi$-periodicity, and results in a half-integer quantization $-i\partial_\theta \rightarrow \ell \in \mathbb{Z}+\frac{1}{2}$. For $\Phi=0$ the rotational symmetry $\ell \leftrightarrow -\ell$ results in doubly degenerate branches. However, by increasing the flux we obtain a single low energy branch ($\ell=\frac{1}{2}$) where the system can be treated as one-dimensional, as long as higher values of $\ell$ are not excited. In the case $\Phi=\frac{1}{2}$, the gap formed due to the nanowire's finite radius is closed and a linear Dirac spectrum is formed. Introducing $\Delta_B$ as the magnetic gap that the Dirac spectrum acquires, we can define the flux in general as $\Phi = \frac{1}{2} - \frac{\Delta_B R}{v}$. Focusing on $\ell=\frac{1}{2}$ we obtain the one-dimensional Hamiltonian 
\begin{equation}
\label{eq: TIN-1D hamiltonian}
\mathcal{H}'_\text{surf} = iv \partial_z\sigma_y + \Delta_B\sigma_z.
\end{equation}
When a spatially varying pairing term $\Delta(z) \psi_{\uparrow}^\dag\psi_{\downarrow}^\dag + \text{h.c.}$ is included \cite{PhysRevB.84.201105_supp}, the nanowire can nucleate Majorana fermions. Here we derive the wave functions of the two Majorana fermions in a single wire with edges at $z=0$ and $z=-L$. For a finite $L$ a small energy splitting occurs due to the hybridization of edge modes, but we will ignore this effect and assume exact zero-modes. To find the zero-modes we solve $H f(z) = 0$ with
\begin{equation}
\label{eq: TIN-ham 1D + sc}
H = (iv\partial_z\sigma_y+\Delta_B\sigma_z)\tau_z - \Delta(z)\sigma_y\tau_y,
\end{equation}
where $\Delta(z)$ has a step-like profile localized around the edges. At the right edge $\Delta(z\gg 0) \rightarrow 0$ and $\Delta(z\ll 0) \rightarrow \Delta$, and the mirror image at the left edge. The magnetic gap $\Delta_B < \Delta$ is kept constant throughout. We will denote the right edge solution by $f(z)$ and the left edge solution by $\bar{f}(z)$. Using the transformation $\tilde{H} = U^\dag H U$ with $U = e^{-i\frac{\pi}{4}(\sigma_y + \tau_y)}$ we obtain  
\begin{equation}
\label{eq: TIN-ham 1D + sc rotated}
\tilde{H} =  \left(\Delta_B\sigma_x   -iv\partial_z\sigma_y\right)\tau_x- \Delta(z)\sigma_y\tau_y.
\end{equation}
This gives us four independent equations of the form $\left(\partial_z - W_j(z)\right)u_j(z)=0$, where $u_j$ are elements of the spinor solving $\tilde{H} g(z) = 0$. In general the solution has the form $u_j(z) = N_0 e^{\int_{z_0}^z W_j(z')dz'}$ with a normalization factor $N_0$. Given the $\Delta(z)$ profile under consideration only one normalizable solution  exists per edge. The appropriate solution at the right edge $(z=0)$ is $u(z) = N_0 e^{\int_{0}^z (\frac{\Delta(z')-\Delta_B}{v})dz'}$, while at the left edge $(z=-L)$ it is $\bar{u}(z) = N_0 e^{-\int_{-L}^z (\frac{\Delta(z')-\Delta_B}{v})dz'}$. In spinor form the solutions are given by
\begin{equation}
\label{eq: TIN-ham 1D - g}
\begin{aligned}
g(z) & = u(z)(0,i,0,0)^T \\
\bar{g}(z) & = \bar{u}(z)(0,0,1,0)^T. \\
\end{aligned}
\end{equation}
We choose a specific realization for $\Delta(z)$ as a step-function and obtain the normalization factor $N_0 = \sqrt{\frac{2\Delta_B(\Delta-\Delta_B)}{v\Delta}}$. Reverting to the original basis we get the spinors
\begin{equation}
\label{eq: TIN-ham 1D - f}
\begin{aligned}
f(z) & = \frac{1}{2}u(z)(i,i,-i,-i)^T \\
\bar{f}(z) & = \frac{1}{2}\bar{u}(z)(1,-1,1,-1)^T , \\
\end{aligned}
\end{equation}
which are used to define the two types of Hermitian Majorana operators $\gamma, \bar{\gamma}$. 
This can be generalized trivially to a Josephson junction by combining two such wires, each hosting a pair: $\gamma_1, \bar{\gamma}_1$ at left side and $\gamma_2, \bar{\gamma}_2$ at the right. 

\section{Effective Field Theory}
\label{eff_app}
Here we show a full derivation of the effective theory by focusing on the low-energy excitations. We first consider the integration of high-energy quasi-particles, and afterwards include the Majorana fermions. This will give us a full description of the bound states in conjunction with the superconducting pairing phase and its dynamics.

\subsubsection{High Energy Quasi-particles}
\label{eff_app-high}
We concentrate on the action describing the superconducting regions of the nanowire given by $S_{\text{sc}}$ in the main text, and their coupling to the weak-link $S_\text{tun}$. Throughout we use $\Delta$ as the largest energy scale to make controlled approximations. In particular we assume that the phase dynamics are slow compared to the superconducting quasi-particles, with the exception of the Majorana fermions. It will be sufficient to examine only one of the junction sides which is governed by the action
\begin{equation}
\label{eq: eff_app-high-action}
S = \int dt dz \left[\Psi^\dag G^{-1}\Psi+\Psi^\dag\eta  + \eta^\dag \Psi\right],
\end{equation}
where $G$ is the Green function defined by $S_{\text{sc}}$. In addition we have defined the auxiliary field $\eta(z,t) = \lambda\delta(z)e^{\frac{i\phi(t)}{2}\tau_z} \mathcal{C}$ and for simplicity omitted the index $j$. Integrating the fermionic fields $\Psi$, $\Psi^\dag$ results in the form $S' = S_0 + S_1$, where
\begin{equation}
\label{eq: eff_app-high-action 0}
S_0 = -\text{tr} \ln(G^{-1}),
\end{equation}
and
\begin{equation}
\label{eq: eff_app-high-action 1}
S_1= \int dt dz \int dt'dz' \eta^\dag(z',t')G(z-z';t,t')\eta(z,t).
\end{equation}
The notation $``\text{tr}"$ designates the trace over the orbital and the Nambu-spin subspaces. 
We start by evaluating the term $S_0$ which describes the slow phase dynamics. The inverse Green function can be written as $G^{-1} = g^{-1} - \chi(t)$, where  $g = \sum_{\omega k} e^{i(kz-\omega t)} g(\omega,k)$ is the bare Green function valid in the mean-field regime, and $\chi(t) = \frac{1}{2}\dot{\phi}\tau_z$ is the deviation from the mean-field. This allows us to recast Eq. (\ref{eq: eff_app-high-action 0})  as $S_0 = \text{tr} \ln(1-g\chi)$, where we ignored the excess term $\text{tr} \ln (g^{-1})$ since it is independent of the phase. An explicit form for $g$ is given by 
\begin{equation}
\label{eq: eff_app-high- g matrix}
g(\omega, k) = \frac{1}{W}
\begin{pmatrix}
A(\Delta_B) & -i P & i U & -D(\Delta_B) \\
i P & A(-\Delta_B) & D(-\Delta_B) & -i U \\
-iU & D(-\Delta_B) & A(-\Delta_B) & i P \\
-D(\Delta_B) & iU & -i P & A(\Delta_B) 
\end{pmatrix},
\end{equation}
with the matrix elements
\begin{equation}
\label{eq: eff_app-high- g matrix elements}
\begin{aligned}
& W = (\Delta_B^2+\Delta^2+v^2k^2-\omega^2)^2 - 4\Delta_B^2\Delta^2 \\
& A(\Delta_B) = \Delta_B(\Delta^2-v^2k^2+\omega^2)-\omega \Delta_B^2-\Delta_B^3 \\
& \qquad \quad - \omega(\Delta^2+v^2k^2-\omega^2) \\
& D(\Delta_B) = \Delta(\Delta^2+v^2k^2 -(\Delta_B+\omega)^2) \\
& P = vk(\Delta_B^2+\Delta^2+v^2k^2-\omega^2) \\
& U = 2vk\Delta_B\Delta.
\end{aligned}
\end{equation}
Since the phase fluctuations are small compared to the energy of the quasi-particles $\sim \Delta$, we can perform a 2nd order expansion in $g \chi$, which gives us
\begin{equation}
\label{eq: eff_app-high- s0}
S_0 \simeq \text{tr}(g\chi)  +\frac{1}{2}\text{tr}(g\chi g \chi).
\end{equation} 
Evaluating the 1st term we get
\begin{equation}
\label{eq: eff_app-high- s0 1st term}
\text{tr}(g\chi) = \text{Tr}\sum_{\omega \omega' k} g(\omega,k)\chi(\omega-\omega')\delta_{\omega \omega'}, 
\end{equation}
where the trace is divided into an orbital sum given by the Fourier components, and a sum over the Nambu-spin subspace denoted by Tr. The Fourier components of the phase fluctuations are given by $\chi(\omega-\omega') =  \frac{i}{2}(\omega-\omega')\phi_{\omega-\omega'}\tau_z$ and with the included delta function $\delta_{\omega \omega'}$ reduce the entire term to zero. The 2nd term in Eq. (\ref{eq: eff_app-high- s0}) has the leading contribution
\begin{equation}
\label{eq: eff_app-high- s0 2nd term 1}
\text{tr} \left(g\chi g \chi\right) = \frac{1}{4}\text{Tr} \sum_{\omega \Omega k} \Omega^2|\phi_\Omega|^2 g(\omega,k)\tau_zg(\omega-\Omega,k)\tau_z.
\end{equation}
We used the replacement $(\omega+\omega')\rightarrow \omega$ and $(\omega-\omega')\rightarrow \Omega$
Note that since $\chi \sim \Omega$ and $g \sim \frac{1}{\Delta}$, the expansion performed in Eq. (\ref{eq: eff_app-high- s0}) is a 2nd order gradient in $\dot{\phi}/\Delta$. In this framework the bare Green function in (\ref{eq: eff_app-high- s0 2nd term 1}) is approximated using $g(\omega-\Omega) \simeq g(\omega)-g'(\omega)\Omega$ which results in
\begin{equation}
\label{eq: eff_app-high- s0 2nd term 2}
\text{tr}(g\chi g\chi) = \frac{1}{4E_C^{(w)}} \left(\sum_\Omega \Omega^2 |\phi_\Omega|^2\right).
\end{equation}
Here we neglected terms proportional to $\Omega^3$, and defined the parameter $E_C^{(w)}$ via
\begin{equation}
\label{eq: eff_app-high- charging energy}
E_C^{(w)} \equiv
\left(\text{Tr} \sum_{\omega k}  g(\omega,k)\tau_zg(\omega,k)\tau_z\right)^{-1}.
\end{equation}
Transforming Eq. (\ref{eq: eff_app-high- s0 2nd term 2}) back to the time-domain we obtain the free part of the phase dynamics resulting from the bulk fermions $S_0 = \int\frac{dt}{8E^{(w)}_C} (\partial_t \phi)^2$. This term can be absorbed into the action of the transmon, given by $S_J$ in the main text, by the replacement $E_C \rightarrow  \frac{E^{(w)}_C E_C}{E^{(w)}_C+E_C}$, accounting for both the capacitance of the Mesoscopic Josephson junction and the nanowire.

We will now proceed to evaluate the second contribution to the action, given by Eq. (\ref{eq: eff_app-high-action 1}). 
We approximate the Green function in Eq. (\ref{eq: eff_app-high-action 1}) as $G \simeq g + g\chi g$ and as a result the action can be split into two terms $S_1 = S_1^{(0)} + S_1^{(\chi)}$. First we consider the case $\chi=0$, in which the Green function can be replaced by the bare diagonal Green function $G(\omega,\omega';k,k') = g(\omega,k)\delta_{\omega \omega'}\delta_{kk'}$, resulting in
\begin{equation}
\label{eq: eff_app-high- S1}
\begin{aligned}
S_1^{(0)} = & \lambda^2\int dt dt' \mathcal{C}^\dag(t') e^{-\frac{i\phi(t')}{2}\tau_z}  \\
& \times\left(\sum_{\omega k} g(\omega,k) e^{-i\omega(t-t')}\right)
e^{\frac{i\phi(t)}{2}\tau_z}\mathcal{C}(t) .
\end{aligned}
\end{equation}
Seeing that we focus on the low energy theory for the $\mathcal{C}$ fermions we can expand the Green function (\ref{eq: eff_app-high- g matrix}) to 
\begin{equation}
\label{eq: eff_app-high- approx g}
\sum_k g(\omega,k) \simeq \sum_k  \left(g(0,k)
+ \left.\omega\frac{\partial g(\omega,k)}{\partial \omega}\right|_{\omega=0} \right).
\end{equation}
To calculate the sum over $k$ we replace it with an integral $\sum_k \rightarrow \frac{1}{\pi v} \int d\varepsilon$, where $\varepsilon = vk$, which results in
\begin{equation}
\label{eq: eff_app-high- sum g0}
g_0\equiv \sum_k g(0,k) = \frac{1}{v}\sigma_y\tau_y,
\end{equation}
and
\begin{equation}
\label{eq: eff_app-high- sum g1}
g_1\equiv
\sum_k \left.\frac{\partial g(\omega,k)}{\partial \omega}\right|_{\omega=0} = \frac{\left(\Delta +2^{3/2}\Delta_B\sigma_x\tau_x \right)}{v(\Delta_B^2-\Delta^2)}.
\end{equation}
Using the identities $\sum_\omega e^{-i\omega(t-t')} = \delta(t-t')$ and $\sum_\omega \omega e^{-i\omega(t-t')} = \delta(t-t') i\partial_t$, Eq. (\ref{eq: eff_app-high- S1}) evaluates to 
\begin{equation}
\label{eq: eff_app-high- S1 1}
\begin{aligned}
S_1^{(0)} = & \lambda^2\int dt  \Big[\mathcal{C}^\dag(t) e^{-\frac{i\phi(t)}{2}\tau_z} \left(g_0-\frac{1}{2}(\partial_t\phi)g_1\tau_z\right)e^{\frac{i\phi(t)}{2}\tau_z}\mathcal{C}(t) \\
& + \mathcal{C}^\dag(t)\left(e^{-\frac{i\phi(t)}{2}\tau_z} g_1e^{\frac{i\phi(t)}{2}\tau_z}\right)
i\partial_t\mathcal{C}(t) 
\Big].
\end{aligned}
\end{equation}
We continue and examine the case $\chi \neq 0$. Contrary to the previous case, the Green function's contribution to $S_1^{(\chi)}$ is of the form $g\chi g$ and therefore not diagonal in $\omega$ and $k$. Explicitly:
\begin{equation}
\label{eq: eff_app-high- S1 chi}
\begin{aligned}
S_1^{(\chi)} = & \lambda^2\int dtdt'  \mathcal{C}^\dag(t') e^{-\frac{i\phi(t')}{2}\tau_z} K(t,t')
e^{\frac{i\phi(t)}{2}\tau_z}\mathcal{C}(t),
\end{aligned}
\end{equation}
where
\begin{equation}
\label{eq: eff_app-high- kernel}
K(t,t') = \sum_{\omega \omega' k} g(\omega,k) \chi(\omega-\omega') g(\omega',k) e^{-i(\omega t - \omega' t')}.
\end{equation}
Since $\chi \sim (\omega-\omega')$, it will be sufficient to take the zeroth order of $g(\omega,k)$ which results in
\begin{equation}
\label{eq: eff_app-high- kernel 2}
K(t,t') \simeq \sum_k g(0,k)\tau_z g(0,k) \sum_{\omega \omega'}\chi(\omega-\omega')e^{-i(\omega t - \omega' t')}.
\end{equation}
By performing the sum over $k$ and using Eq. (\ref{eq: eff_app-high- g matrix}) we can show that $K(t,t') =0$. Thus the $S_1^{(0)}$ term given by Eq. (\ref{eq: eff_app-high- S1 1}) is the only contribution to $S_1$. 
Finally, combining both sides of the junction and adding $S_{_W}$, we get the full action for the weak-link fermions:
\begin{equation}
\label{eq: eff_app-high- c action}
\begin{aligned}
& S_\mathcal{C}   =  \int dt  \mathcal{C}^\dag \Big[ \mathcal{J}i\partial_t -(\varepsilon + \Delta_B\sigma_z)\tau_z \\
&  + \sum_{j=1,2} \left(\Gamma_j \sigma_y\tau_y e^{i\phi_j(t)\tau_z} - \alpha_j \dot{\phi}_j \tau_z \right)
\Big]\mathcal{C} -U c^\dag_\uparrow c_\uparrow c^\dag_\downarrow c_\downarrow, \\
\end{aligned}
\end{equation}
where
\begin{equation}
\label{eq: eff_app-high- J}
\mathcal{J} = 1 + \sum_{j=1,2} \lambda^2_j e^{-\frac{i\phi_j(t)}{2}\tau_z} g_1e^{\frac{i\phi_j(t)}{2}\tau_z}.
\end{equation}
We introduced $\Gamma_j = \lambda_j^2/v$ which serves as the induced pairing strength. Since the integration systematically removes high energy excitations it should satisfy $\Gamma_j \ll \Delta$. Due to the integration process a fermion-phase interaction term appears with coupling constant
 $\alpha_j =\frac{\lambda_j^2 \Delta}{2v(\Delta^2-\Delta_B^2)}$. This term is meaningful as long as $E_C$ and $E_C'$ remain finite, otherwise it can be gauged out, similar to a vector potential of a massless particle.

\subsubsection{Majorana Fermions}
A general quasi-particle $\Psi(z)$ in the superconductor can be defined as a combination of the Bogoliubov operators, with two of them given by the Majorana fermions and the rest are high energy bulk states
\begin{equation}
\label{eq: eff_app-maj - psi}
\Psi(z) = \frac{iu(z)}{\sqrt{2}}\tau_z \gamma + \frac{\bar{u}(z)}{\sqrt{2}}\sigma_z \bar{\gamma} + \sum_n U_n(z)\Gamma_n.
\end{equation}
We have used the results and notation of Eq. \ref{eq: TIN-ham 1D - f} to define the contribution of the Majorana fermions. 
The operators $\Gamma_n$ are spinors of Bogoliubov quasi-particles with the amplitudes in matrix form $U_n(x)$. In the previous section we used momentum states as eigenstates of a one-dimensional superconducting wire. This approach can be justified here as well with the approximation $\sum_n U_n(z) \Gamma_n \simeq \sum_k e^{ikz}\Psi_k$, as the Majorana fermions are zero modes energetically isolated from the gapped quasi-particles.
We can project Eq. (1) in the main text to the Majorana sector by the replacement $\Psi_1(z) \rightarrow \frac{iu(z)}{\sqrt{2}}\tau_z\gamma_1$ and $\Psi_2(z) \rightarrow \frac{\bar{u}(z)}{\sqrt{2}}\sigma_z\bar{\gamma}_2$, while the rest of the quasi-particles integrated. Thus we obtain
\begin{equation}
\label{eq: eff_app-maj - s_gamma}
\begin{aligned}
S_{\gamma} & = 
\int dt \Big(\gamma_1 i\partial_t \gamma_1 + \bar{\gamma}_2 i\partial_t \bar{\gamma}_2  \\
 + & \frac{iw_1}{\sqrt{2}}\gamma_1e^{\frac{i\phi_1(t)}{2}\tau_z}\tau_z \mathcal{C}+\frac{w_2}{\sqrt{2}}\bar{\gamma}_2e^{\frac{i\phi_2(t)}{2}\tau_z}\sigma_z \mathcal{C} + \text{h.c.}\Big),
\end{aligned}
\end{equation}
where $w_1 = u(0)\lambda_1$ and $w_2 = \bar{u}(0)\lambda_2$.
Note that we neglected the Majorana fermions on the outermost edges of the junction since their hybridization is exponentially small in the length of the nanowire. All the low-energy contributions given by Eq.(\ref{eq: eff_app-high- c action}), (\ref{eq: eff_app-maj - s_gamma}) and $S_J$ in the main text are combined into an effective action $S_\text{eff}$, which is used in the next section to derive the effective Hamiltonian.

\section{Derivation of the Hamiltonian}
\label{ham}
To derive the Hamiltonian we extract the canonical variables from the Lagrangian in $S_\text{eff} = \int dt \,\mathcal{L}_\text{eff}$, written in terms of the phase difference $\varphi = \phi_1-\phi_2$ and the average phase $\delta = (\phi_1+\phi_2)/2$.
We define $\mathcal{P}_{X}=\frac{\partial \mathcal{L}_\text{eff}}{\partial (\partial_t X)}$ as the conjugate momentum to  $X = \left\{\varphi, \delta,\gamma_1,\bar{\gamma}_2, \mathcal{C}\right\}$ and 
by employing Legendre's transformation we obtain the Hamiltonian in the form 
\begin{equation}
\label{eq: ham - legendre}
H_\text{eff}= \sum_X \mathcal{P}_X (\partial_t X) - \mathcal{L}_\text{eff} = H_T + H_\mathcal{C} + H_\gamma.
\end{equation}
Since the dynamics of the phase $\partial_t \phi_j$ dictate the charge fluctuations, we identify $\mathcal{P}_\varphi $ with the relative number of cooper pairs between the two superconducting islands, and $\mathcal{P}_\delta$ with the total number of cooper pairs  which exceed neutrality in the islands. Explicitly these are 
\begin{equation}
\label{eq: ham - n and N definitions}
\begin{aligned}
\mathcal{P}_\varphi  & =  \frac{\partial_t \varphi(t)}{8E_C} - \alpha_- \mathcal{C}^\dag \tau_z\mathcal{C} \\
\mathcal{P}_\delta  & = \frac{\partial_t \delta(t)}{2E_C'} - \alpha_+ \mathcal{C}^\dag \tau_z\mathcal{C},
\end{aligned}
\end{equation}
where $\alpha_- = \frac{1}{2}(\alpha_1 - \alpha_2)$ and $\alpha_+ = \alpha_1 + \alpha_2$. With these definitions we extract $H_J$ from $S_\text{eff}$ as
\begin{equation}
\label{eq: ham-josephson}
\begin{aligned}
H_T & = 4E_C\left(\hat{n} +\alpha_-\mathcal{C}^\dag \tau_z \mathcal{C}\right)^2 \\
& + E_C'\left(\hat{N} +\alpha_+\mathcal{C}^\dag \tau_z \mathcal{C}\right)^2 -E_J \cos(\hat{\varphi}).
\end{aligned}
\end{equation}
We used the quantized versions of the conjugate momenta: $\mathcal{P}_\varphi \rightarrow \hat{n}$ and $\mathcal{P}_\delta \rightarrow \hat{N}$.
Note that in the symmetric regime $(\lambda_1 = \lambda_2)$, the one used in the main text, the coupling constant $\alpha_-$ vanishes. Similarly to Eq. (\ref{eq: ham-josephson}), we get the Hamiltonian for the weak-link fermions together with the transmon
\begin{equation}
\label{eq: ham-weak}
\begin{aligned}
H_\mathcal{C} & = \mathcal{C}^\dag  \Big[ \left(\varepsilon + \Delta_B\sigma_z\right)\tau_z  
+ e^{-i\frac{\hat{\delta}}{2}\tau_z}
\Big((\Gamma_1+\Gamma_2)\cos\left(\frac{\hat{\varphi}}{2}\right)\tau_y \\
& -(\Gamma_1-\Gamma_2)\sin\left(\frac{\hat{\varphi}}{2}\right)\tau_x\Big)
e^{i\frac{\hat{\delta}}{2}\tau_z}\sigma_y\Big]\mathcal{C} +U c^\dag_\uparrow c_\uparrow c^\dag_\downarrow c_\downarrow.
\end{aligned}
\end{equation}
The coupling of the weak-link fermions to the Majorana fermions is given by
\begin{equation}
H_\gamma = 
\frac{1}{\sqrt{2}}\left[
-iw_1\gamma_1e^{\frac{i(2\hat{\delta}+\hat{\varphi})}{4}\tau_z}\tau_z -w_2\bar{\gamma}_2e^{\frac{i(2\hat{\delta}-\hat{\varphi})}{4}\tau_z}\sigma_z \right] \mathcal{C}+\text{h.c.}
\end{equation}

\section{Highly Transmitting Limit for the Weak-Link}
\label{continuous}
Here we look at an alternative approach for an effective theory of Andreev bound states. One which does not include a tunneling process between the superconducting islands and the weak-link and is based solely on Andreev scattering processes between the superconducting and the normal regions of the nanowire with $\Delta_B=0$. The normal region is simply given by
\begin{equation}
\label{eq: continuous - sw}
S_{_W} = \int dt \int dz \chi^\dag \left(i\p_t + iv\p_z\sigma_y\tau_z\right)\chi, 
\end{equation}
with $\chi=\frac{1}{\sqrt{2}}(\chi_\uparrow, \chi_\downarrow, \chi^\dag_\uparrow, \chi^\dag_\downarrow)^T$. As in the main text, a contribution of the superconducting regions also exists in the form of the fields $\Psi_j$ ($j=1,2$). Without loss of generality, to develop the effective action, we focus on one side of the weak link and omit the index $j$ for simplicity. In addition, we introduce a Lagrange multiplier $\zeta(t)$, $\zeta^\dag(t)$ via
\begin{equation}
\label{eq: continuous - s_zeta}
S_{\zeta} = \int dt \left[\zeta^\dag(t)\left(\chi(0,t) - e^{-i\frac{\phi(t)}{2}\tau_z}\Psi(0,t)\right) +\text{h.c.} \right], 
\end{equation}
which insures the continuity of the fields between the superconducting and the normal regions. Similarly to the derivation of the low-energy effective action in the previous section, we start by integrating out the $\Psi$ fields to obtain
\begin{equation}
\label{eq: continuous - s_zeta_eff}
\begin{aligned}
S_{\zeta} &= \int dt \Big[\left( \int dt' \zeta^\dag(t')  e^{-i\frac{\phi(t')}{2}\tau_z} g(t-t') e^{i\frac{\phi(t)}{2}\tau_z} \zeta(t)\right) \\
& + \zeta^\dag(t)\chi(0,t)+\chi^\dag(0,t)\zeta(t)\Big].
\end{aligned}
\end{equation}
Since we already developed the action for the phase fluctuations with the expansion in Eq. (\ref{eq: eff_app-high- s0}), we leave out the $\text{tr} \ln (G^{-1})$ term from this section and focus only on the dynamics of $\chi$. The green function in Eq. (\ref{eq: continuous - s_zeta_eff}) is given by $g(t-t') = \sum_{\omega k} g(\omega,k) e^{-i\omega(t-t')}$, with $g(\omega,k)$ defined in (\ref{eq: eff_app-high- g matrix}). To find an approximate form for $g(t-t')$ we use the results of Eqs. (\ref{eq: eff_app-high- approx g}),(\ref{eq: eff_app-high- sum g0}) and (\ref{eq: eff_app-high- sum g1}) with $\Delta_B=0$. This gives us
\begin{equation}
g(t-t') \simeq \frac{\delta(t-t')}{v}\left(\sigma_y\tau_y - \frac{i\p_t}{\Delta}\right).
\end{equation}
We proceed and integrate out the Lagrange Multipliers from Eq. (\ref{eq: continuous - s_zeta_eff}), giving us a pairing term localized at the edge of the weak-link, originating in $g^{-1}(t-t')$. Combining this pairing contribution from both sides of the weak-link with Eq. (\ref{eq: continuous - sw}), we obtain the effective action for $\chi$ in the form
\begin{equation}
\label{eq: continuous s_chi}
\begin{aligned}
S_{\chi} & = \int dt \int dz \; \chi^\dag \Big[\mathcal{J}_\chi i\p_t + iv\p_z \sigma_y\tau_z \\
& + \sum_{j=1,2} \delta(z-z_j)\left(v\sigma_y\tau_y e^{i\phi_j(t) \tau_z} -\frac{v}{2\Delta}\dot{\phi}_j \tau_z \right) \Big]\chi,
\end{aligned}
\end{equation}
where
\begin{equation}
\mathcal{J}_\chi = 1 + \frac{v}{\Delta}\sum_{j=1,2} \delta(z-z_j).
\end{equation}
Here $z_1 = -L/2$ and $z_2=L/2$ are the edges of the weak-link.

We would like to stress the similarity between this result, and the one obtained in Eq. (\ref{eq: eff_app-high- c action}) using a tunneling term $S_\text{tun}$. Both results showcase similar structure and coupling between the microscopic and macroscopic degrees of freedom. Specifically the induced pairing and the coupling of $\dot{\phi}_j$ to the charge density near the appropriate edge. Note that the tunneling parameter $\lambda$ which determines the energy scale of the Andreev bound state $\Gamma = \lambda^2/v$ does not appear in this continuous model. Instead, the energy of the Andreev bound states here is determined by $\sim v/L$, which we take to be small compared to $\Delta$. 

The full action of the system includes also the contribution of the transmon $S_J$, given in the main text. By adding the action of the transmon, and changing the phases to $\varphi$ and $\delta$, we can define the Hamiltonian of the system using the transformation $H = \frac{\p \mathcal{L}}{\p \dot{\chi}}\dot{\chi} + \hat{n}\dot{\varphi} + \hat{N}\dot{\delta} - \mathcal{L}$. Here $\mathcal{L}$ is the Lagrangian of the full action $S = S_J + S_{\chi}$. The generalized momenta are defined as
\begin{equation}
\begin{aligned}
\hat{n} & = \frac{\p \mathcal{L}}{\p \dot{\varphi}} = \frac{\dot{\varphi}}{8E_C} - \frac{v}{4\Delta}
\int dz\; \chi^\dag (\delta(z-z_1) - \delta(z-z_2))\tau_z\chi, \\
\hat{N} & = \frac{\p \mathcal{L}}{\p \dot{\delta}} = \frac{\dot{\delta}}{2E_C'} -\frac{v}{2\Delta}
\int dz\; \chi^\dag (\delta(z-z_1) + \delta(z-z_2))\tau_z\chi.
\end{aligned}
\end{equation}
These give us the Hamiltonian:
\begin{equation}
\begin{aligned}
H & = 4E_C \left(\hat{n} + \frac{v}{4\Delta}
\int dz \chi^\dag (\delta(z-z_1) - \delta(z-z_2))\tau_z\chi\right)^2
 \\
& + E_C' \left(\hat{N} + \frac{v}{2\Delta}\int dz \chi^\dag (\delta(z-z_1) + \delta(z-z_2))\tau_z\chi\right)^2 \\
&-E_J \cos(\hat{\varphi}) 
+ \int dz \chi^\dag \Big[-iv\p_z\sigma_y\tau_z \\
& -v \sigma_y\tau_y\left(e^{i\frac{\hat{\varphi}}{2}\tau_z}\delta(z-z_1) 
+ e^{-i\frac{\hat{\varphi}}{2}\tau_z}\delta(z-z_2)\right)e^{i\hat{\delta}\tau_z}\Big]\chi.
\end{aligned}
\end{equation}
Here we see that the operator $\hat{n}$ couples to the relative charge between the edges of the weak-link, and $\hat{N}$ couples to the total charge at the edges.

%

\end{document}